\documentstyle[preprint,aps]{revtex}
\input{psfig}

\begin{document}

\title{Motion of buoyant particles and coarsening
of solid-liquid mixtures in a random acceleration field}
\author{J. Ross Thomson$^{1}$, Jaume Casademunt$^{2}$, Fran\c{c}ois
Drolet$^{1}$ and Jorge Vi\~nals$^{1,3}$}
\address{$^{1}$ Supercomputer Computations Research Institute, Florida State
University, Tallahassee, Florida 32306-4052; $^{2}$ Departament
d'Estructura i Constituents de la Materia, Universitat de Barcelona,
08028 Barcelona, Spain; $^{3}$ Department of Chemical Engineering,
FAMU/FSU College of Engineering, Tallahassee, Florida 32310}

\date{\today}

\maketitle

\begin{abstract}

Flow induced by a random acceleration field (g-jitter) is
considered in two related situations that are of interest for microgravity
fluid experiments: the random motion of an isolated buoyant particle and
coarsening of a solid-liquid mixture.
We start by analyzing in detail actual accelerometer data gathered during a 
recent microgravity mission, and obtain the values of the parameters defining a 
previously introduced stochastic model of this acceleration field. 
We then study the motion
of a solid particle suspended in an incompressible fluid that is
subjected to such random accelerations. The displacement of the particle
is shown to have a diffusive
component if the correlation time of the stochastic acceleration is finite 
or zero,
and mean squared velocities and effective diffusion coefficients are obtained
explicitly. Finally, the effect of g-jitter on coarsening of a
solid-liquid mixture is considered.
Corrections due to the induced fluid motion are calculated, and estimates
are given for coarsening of Sn-rich particles in a Sn-Pb eutectic
fluid, experiment to be conducted in microgravity in the near future.

\end{abstract}


\narrowtext

\section{Introduction}
\label{sec:intro}

With recent frequent access to a microgravity environment, more
attention is being paid to a precise characterization of the
effective acceleration environment onboard spacecraft (g-jitter), 
as well as to the analysis of potential effects of such an environment on a
number of experiments, compared to an ideal zero gravity situation
\cite{re:walter87,re:koster90,re:rath92,re:antar93}. We study in this
paper the motion induced on particles that are suspended in an incompressible
fluid by an external random acceleration field.
As an extension, coarsening of a solid-liquid mixture is considered, and
the effects of g-jitter estimated for the case of a Sn-Pb eutectic. This
system will be studied in microgravity in the near future.

Whereas qualitative information on the residual acceleration field
onboard spacecraft has been available for some time, it is only 
recently that a systematic effort has been made to collect long temporal
sequences of acceleration data over a fairly wide frequency range 
\cite{re:martin92,re:alexander90}. In
the frequency range we study ($10^{-1} - 10^{2}$ Hz), the
SAMS project now routinely determines for each mission the three components
of the residual acceleration field at selected points in the spacecraft.
This includes, in some cases, sensor heads at the same location where an
experiment potentially susceptible to this kind of disturbances is being
conducted.

From a theoretical point of view, the first issue to be addressed 
concerns the introduction of a suitable model of the residual acceleration 
field. Based on available accelerometer data and their associated power
spectra, it seems necessary to distinguish
between frequency components that can be modeled as systematic
(or deterministic), and thus traced back to some mechanical device
producing a periodic disturbance of known amplitude and frequency, and
random components arising from a number of independent sources with variable
frequencies and intensities. We further note that physical sources of
accelerations contribute to the overall acceleration environment in two ways: 
directly, and indirectly by exciting some of the natural vibration modes of the
spacecraft.

Most of the studies to date have focused on a deterministic
acceleration field modeled as a superposition of periodic functions
of fixed amplitudes and frequencies 
\cite{re:kamotani81,re:jacqmin90,re:alexander91,re:jue92,re:farooq94}. 
Also, some studies have considered the effects of short and isolated pulses
\cite{re:alexander91}. The approach presented in
this paper, on the other hand, models g-jitter as a random process in time
\cite{re:zhang93,re:thomson95}. We assume
that the process obeys Gaussian statistics, consistent with the
assumption that many independent sources contribute to the acceleration
field, thus requiring the knowledge of
only the first two statistical moments of the effective acceleration
field $\vec{g(t)}$. We choose
$\left< \vec{g} \right> = 0$, where $\left < ~ \right >$ denotes an
ensemble average. In general, a nonzero average can be incorporated
into a steady component, and $\vec{g}$ redefined to be the deviation
from the average. We do not consider here the effect of this component.
To model the statistical behavior of the second moment we chose
a Gaussian process called narrow band noise, defined by the correlation
function,
\begin{equation} 
\label{eq:nbnt}
C_{ij}(t-t') = \left< g_{i}(t) g_{j}(t') \right> = \delta_{ij}
\left< g^{2} \right> e^{- |t-t'|/\tau} \cos{\Omega (t-t')},
\end{equation} 
where $g_{i}(t)$ is any of the three components of the acceleration
field, and $\left< g^{2} \right>$, $\tau$ and $\Omega$ are three
constants which characterize the process: its intensity, a correlation
time and a characteristic angular frequency. A particular advantage of
this process is that it allows interpolation between two well known
limits: the white noise limit when $\Omega \tau \rightarrow 0$ with
$\left< g^{2} \right> \tau = D$ finite, in which no frequency component
is preferred, and monochromatic noise when $\Omega \tau \rightarrow
\infty$, $\left< g^{2} \right>$ finite, in which each realization of
the noise is a periodic function of angular frequency $\Omega$. In this
case, the ensemble refers to a distribution of amplitudes and phases,
with identical angular frequency for each realization. Monochromatic
noise is akin to the deterministic studies in which $\vec{g}(t)$ is
modeled by a periodic function, but still retains random values of the
amplitude and phase.

In Section \ref{sec:g-jitter}, we present a statistical analysis of a long
time series gathered by the SAMS team during the recent SL-J mission. A
window of approximately six hours (sampled at 250 Hz) is analyzed
to determine the existence of deterministic and random components,
and to calculate the values of the parameters needed to characterize
both. A given time series can appear to be
deterministic or stochastic depending on the range analyzed: 
If a random function is correlated over times of the
order of $\tau$, its time series will appear deterministic when
analyzed over time windows $T \ll \tau$ and random otherwise.  In the
case of the SAMS time series that we have analyzed, we find
that there is a systematic or deterministic
component of frequency $17$ Hz. The rest of the spectrum is comprised of
a superposition of random components with
small correlation time, and a white noise background (with a
correlation time no longer than the sampling period of $1/250~s$.
We also find significant deviations from Gaussianity, mainly in
the larger amplitude impulses. It may be necessary to introduce other
stochastic models that are not Gaussian to study these contributions
(shot noise, for example).

Section \ref{sec:random_walk} considers the motion of a particle suspended 
in an incompressible fluid of different density, when the fluid is
subjected to the acceleration field described in Section \ref{sec:g-jitter}. 
If the residual acceleration field is deterministic and periodic, the
suspended particle performs an oscillatory motion, with both
velocity and displacement bounded. If, on the other hand,
the acceleration field is random, the mean squared velocity of the
particle is bounded, but its mean squared displacement grows linearly in time. 
The effective diffusive coefficient is obtained as a function of the
parameters of the fluid and noise. In particular, we note that
measurement of the mean squared displacement of a suspended particle in
a microgravity environment would provide an independent data set from which 
one could infer the parameters that characterize the residual
acceleration field. A similar principle has been used to design a passive
accelerometer system \cite{re:matisak94} to obtain the steady component
of the residual acceleration field from the linear drift of a suspended
particle. Not surprisingly, it has
proved difficult in actual microgravity conditions to maintain a well defined 
alignment of the container with respect to the residual acceleration field.

Section \ref{sec:coarsening} addresses the effects of g-jitter on
coarsening of a solid-liquid mixture. A random acceleration field induces
a random velocity field that may lead to enhanced 
coalescence and solute transport. The analysis focuses on 
the solid-liquid mixture Pb-Sn which will be studied in microgravity in the 
near future. We find that g-jitter effects are small for the
conditions of the experiment, and with the values of the noise found in
Section \ref{sec:g-jitter}.

\section{Time series analysis of g-jitter during the SL-J mission}
\label{sec:g-jitter}

The analysis described in this section is based on actual g-jitter data
collected during the SL-J mission (SAMS-258), that flew on September
13--20, 1993. We
have focused on the head A SAMS detector, and studied the
series during the time window MET 0017 to MET 0023, roughly a period of
six hours. All three Cartesian components of the residual acceleration
field have been included in the analysis. The sampling frequency is $250
$ Hz.
%
%
The data used was gathered continuously throughout the period mentioned,
with automatic re-calibration of the sensor heads when needed
(corrections for the signal gain have been taken into account according
to the calibration data also gathered during the mission).

We do not focus here on some basic statistical properties of the signal
which are already automatically monitored (its running mean and root
mean square values), but address two basic points: (a) the existence of
deterministic and random components during this particular
observation period, and (b) the Gaussian nature of the time series.

Consider first a temporal series $g(t)$ and its power spectrum over a
finite window $[ -T, T ]$ defined by
\begin{equation}
\label{eq:pndef}
P_{T}(n) = \frac{1}{2T} \int_{-T}^{T} C(t) e^{-i\frac{n \pi t}{T}} dt,
\end{equation}
where $C(t)$ is the autocorrelation function defined in Eq.
(\ref{eq:nbnt}). The autocorrelation function can be obtained from
$P_{T}(n)$ as a Fourier series,
\begin{equation}
C_{T}(t) = \sum_{n = - \infty}^{\infty} P_{T}(n) e^{i\frac{n \pi t}{T}},
\end{equation}
where we have introduced the notation $C_{T}(t)$ to indicate that
$C_{T}(t) = C(t)$ in $[ -T, T ]$, and is periodic outside of this
interval. The integral in Eq. (\ref{eq:pndef}) can be evaluated
explicitly to yield,
\begin{eqnarray}
\label{eq:pn}
P_{T}(n) = \frac{<g^{2}>}{2T} \left[ 
\frac{ e^{-\lambda T} \left( - \lambda \cos ( \frac{n \pi}{T} -
\Omega)T + ( \frac{n \pi}{T} - \Omega) \sin ( \frac{n \pi}{T} -
\Omega)T \right) + \lambda
}{
\lambda^{2} + \left( \frac{n \pi}{T} - \Omega \right)^{2}
} + \right. \nonumber \\
\left.  \frac{ e^{-\lambda T} \left( - \lambda \cos ( \frac{n \pi}{T} +
\Omega)T + ( \frac{n \pi}{T} + \Omega) \sin ( \frac{n \pi}{T} +
\Omega)T \right) + \lambda 
}{
\lambda^{2} + \left( \frac{n \pi}{T} + \Omega \right)^{2}
} \right] ,
\end{eqnarray}
where $\lambda=1/\tau$. 
In the white noise limit $P_{T}(n) = D/T$, whereas in the monochromatic
limit $P_{T}(n) = \frac{<g^{2}>}{2} \{ \delta_{n\pi /T, \Omega} +
\delta_{n\pi /T, - \Omega} \}$. In the first case, the dominant
contribution comes from the term $1/\tau$ in Eq. (\ref{eq:pn}), whereas
in the second case it comes from the term proportional to $\sin (
\frac{n \pi}{T} - \Omega)T $. The important point to notice is that in
the white noise limit $P_{T}(n)$ is inversely proportional to the window
size $T$, whereas in the monochromatic limit, $P_{T}(n)$ is independent
of $T$. Thus, we argue, an
analysis of the power spectrum $P_{T}(n)$ as a function of the
window size can provide information on the existence of deterministic
or random contribution, at least within the available ranges of $T$ and
$\tau$. 

These results are in fact more generally valid and not restricted to
narrow band noise. Consider the integral,
\begin{equation}
\label{eq:clt}
\hat{g}_{T} (n) = \frac{1}{2 T} \int_{-T}^{T} dt e^{-i \frac{n \pi t}{T}} g(t).
\end{equation}
If $g(t)$ is a random process, with a correlation time $\tau \ll T$,
then for each $n$ $\hat{g}_{T}(n)$ is the sum of
approximately $2T/\tau$ statistically independent variables. Therefore,
according to the Central Limit Theorem, the integral will obey
Gaussian statistics, with variance ${\cal O}(T/\tau)$. As a
consequence, $\hat{g}_{T}(n) \sim {\cal O}(\frac{1}{T} \sqrt{T/\tau})
= {\cal O}(\frac{1}{\sqrt{T}})$ or $P_{T}(n) = | \hat{g}_{T}(n)
|^{2} \sim {\cal O}(\frac{1}{T})$, in
agreement with the result obtained for narrow band noise in the limit of
short correlation time. On the other
hand, for most deterministic functions $\hat{g}_{T}(n) \sim
{\cal O}(1)$ instead, and $P_{T}(n)$ is independent of the window size
$T$.

\begin{figure}
\vspace{1cm}
\psfig{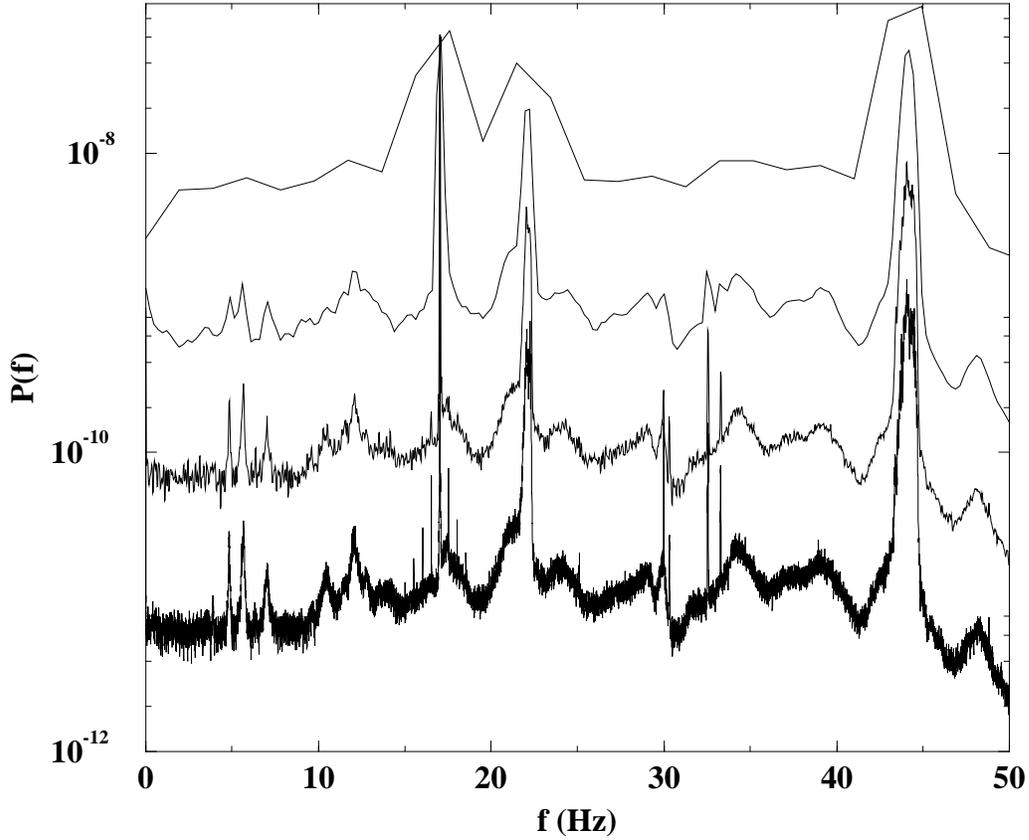}
\caption{Power spectrum as a function of frequency averaged over
a six hour interval during the SL-J mission. The curves shown correspond to
spectra calculated over windows of size (from top to bottom)
$N = 64, 512, 4096$ and 32768. The amplitude of the peak at $f = 17~Hz$ is 
independent of $N$, whereas the amplitude of the peaks at both $22~Hz$ and 
$44~Hz$ decreases with $N$. Also in this latter case, the shape of the peaks 
is independent of $N$.}
\label{fi:power-spectrum}
\end{figure}

We have obtained an estimate of $P_{T}(n)$ for the time
series of $g(t)$ obtained during the SL-J mission and for a range of values 
of $T$. Since the time series is discrete, we consider windows comprising $N$
data points, with $N \Delta t = 2T$ where $1/\Delta t = 250 s^{-1}$ is
the sampling rate (further details on various methods to estimate power spectra
can be found in ref. \cite{re:priestley81}).
Briefly, the power spectrum for a stationary process, one in which its
statistical properties are independent of time, is calculated by averaging 
$P_{T}$, also known as the periodogram. 
The relative statistical error associated with a single
periodogram is 100\% for all frequencies. Reduction in error by a factor
proportional to $1/N_{p}$ can be achieved by averaging $N_{p}$ periodograms 
calculated over disjoint time intervals. The estimate of the power
spectrum presented here is obtained from approximately 6 hours of data,
sampled at 250 Hz. Each periodogram is calculated for a fixed number
of sample points (beginning with 64 and increasing by factors of two)
and then averaged over the entire 6 hour period. The resulting estimates of
the power spectra are summarized in Fig. \ref{fi:power-spectrum}. 
The power spectrum is broadband, with a few peaks at fixed frequencies. 
The background intensity does decrease with increasing $N$, indicating its 
random nature. 

\begin{figure}
\psfig{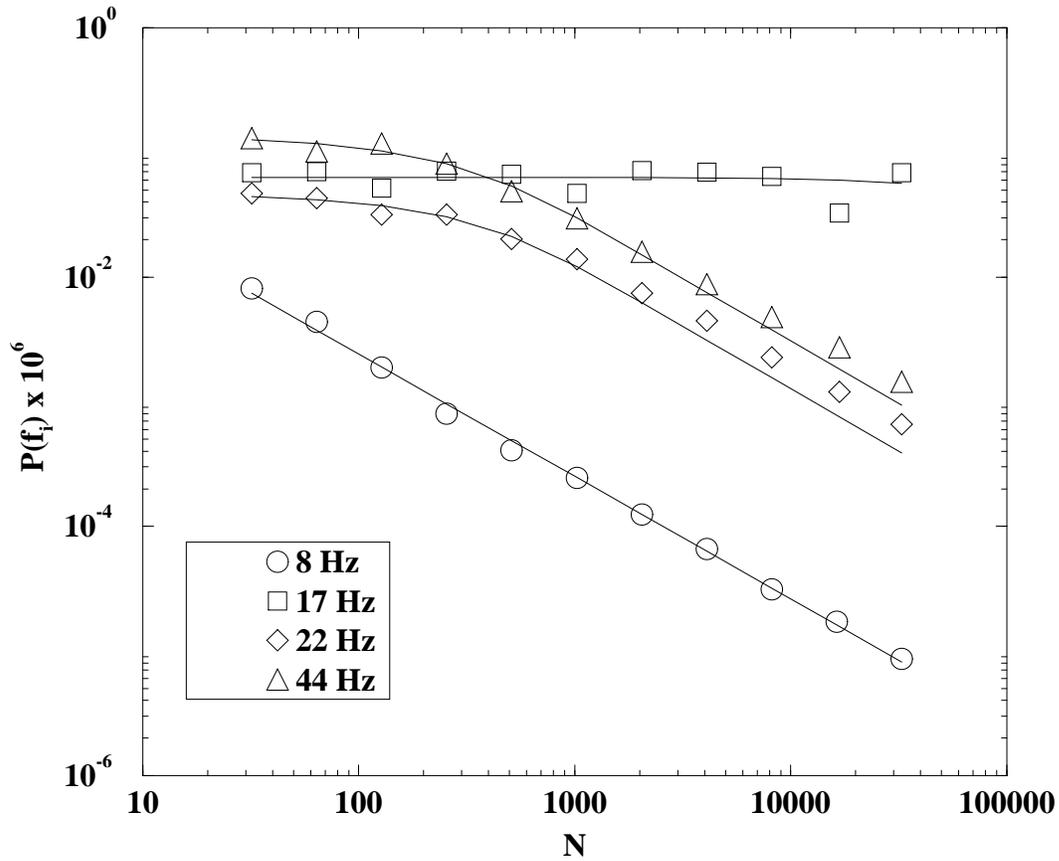}
\caption{Amplitude versus window size $N$ for a few selected frequencies
to display their deterministic or random nature. The amplitude of the
$17~Hz$ component remains independent of $N$ indicating its deterministic
character for the range of window sizes analyzed. Two other components
display mixed behavior, with a finite correlation time of the order of $1~s$.
There is also a clear white noise background, exemplified by the 
amplitude of the power spectrum at $8~Hz$.}
\label{fi:amplitudes}
\end{figure}

To further elucidate the scaling with $T$, we show in Fig.
\ref{fi:amplitudes} the value of $P_{T}$ at selected frequencies as a
function of $T$. The frequency components shown in this figure include
the peaks of Fig. \ref{fi:power-spectrum}, and one intermediate value.
Three types of behavior emerge. First, the value of $P_{T}(f = 17)$ Hz
is independent of $T$ for the range of window lengths studied.
Therefore, and within this range, this component appears to be
deterministic in nature with an amplitude $\sqrt{ < g^{2} >} = 3.56
\times 10^{-4} g_{E}$, where $g_{E}$ is the intensity of the
gravitational field on the Earth's surface. There are two additional
components that have a finite correlation time. We have fitted the
amplitude of the peak to $\left< g^{2} \right> \tau \left( 1 -
e^{-T/\tau} \right)/2T$ and estimated for the component at 22
Hz $\sqrt{ < g^{2} >} = 3.06 \times 10^{-4} g_{e}$ and
$\tau = 1.09 s$, whereas for $44 Hz$ we find $\sqrt{ < g^{2} >} = 5.20
\times 10^{-4} g_{E}$ and $\tau = 0.91 s$. As an estimate of the white
noise background, we obtain from the slope of the intensity of the $8
Hz$ component versus $N$ the value $D = 8.61 \times 10^{-4}
cm^{2}/s^{3}$.

In summary, assuming that the various frequency components can be
studied independently, and that they are independent of the broadband
background, we conclude that the time series analyzed contains a
deterministic component (i.e., a component with a correlation time
larger than the largest window studied), a few isolated components of
large amplitude but small correlation time, and a fairly constant
background, of smaller amplitude, and very small correlation time.

\begin{figure}
\psfig{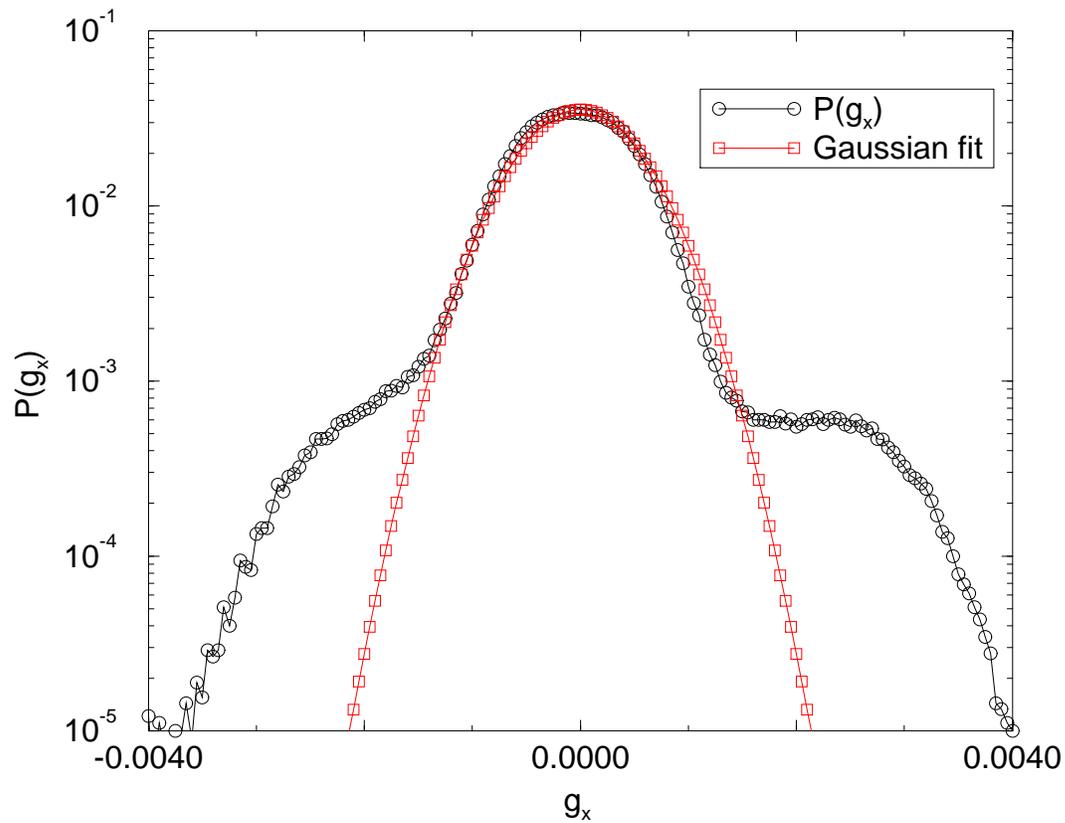}
\caption{Histogram of amplitudes of the residual acceleration along a
particular direction, and a fit to a Gaussian distribution. The distribution
is nearly Gaussian for small values of $g$, but deviates significantly
near the wings. Values of the gravitational field intensity are given
relative to $g_{E}$.}
\label{fi:gaussian}
\end{figure}

To further investigate the statistical nature of the acceleration we
calculate a number of statistical moments of $g(t)$. 
We first present the (one-point) probability distribution of $g$ obtained 
from a histogram of the time series. The histogram comprises 200 bins of width
$0.008 g_{E}$. The result is shown in Fig. (\ref{fi:gaussian}), together
with a fit to a Gaussian distribution. It is apparent that the
distribution is substantially of Gaussian form at low amplitudes, but there are
significant deviations near the tails. We only show in the figure the
$x$ component of the acceleration field. The distribution for the other
two components is virtually identical.

\begin{figure}
\psfig{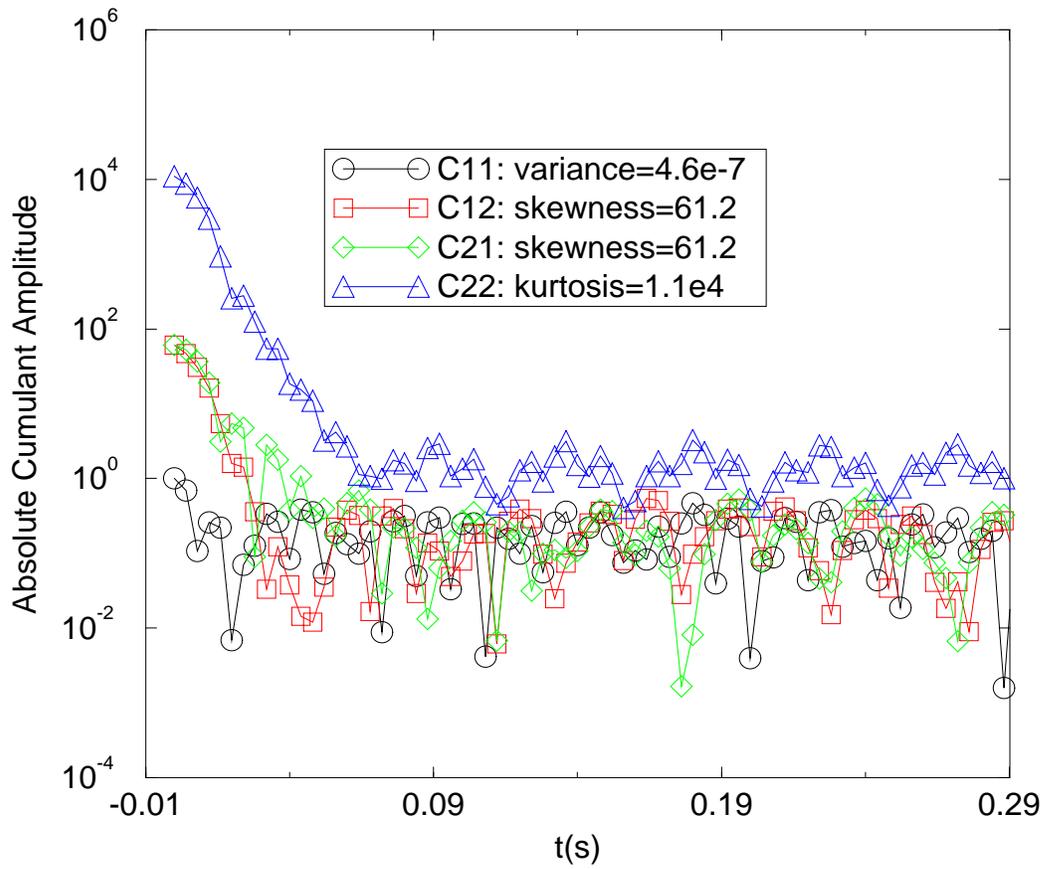}
\caption{Higher order moments of the distribution of the residual
acceleration showing deviations from Gaussianity.}
\label{fi:moments}
\end{figure}

Figure (\ref{fi:moments}) presents the results of higher statistical
moments. Normalized cumulants have been introduced as follows:
\begin{equation}
\label{eq:cumulants}
C_{mn}(t) = \frac{<<g(0)^{m}g(t)^{n}>>}{<<g^{2}>>^{(m+n)/2}},
\end{equation}
where $<< \ldots >>$ is the standard cumulant \cite{re:ma85}.
For a Gaussian process, all cumulants should be zero except for 
$C_{20}(t)=C_{02}(t)$ and $C_{11}(t)$.
Also note that $C_{12}(0) = C_{21}(0)$ reduces to the standard
definition of skewness of a distribution, and $C_{22}(0)$ to its
kurtosis. Again, significant deviations from Gaussianity are found.
Further analysis is needed to elucidate whether the deviations from
Gaussianity in both Figs. (\ref{fi:gaussian}) and (\ref{fi:moments})
originate entirely from the deterministic component at 17 Hz, or are a
more intrinsic feature of the random components.

\section{Motion of a buoyant particle}
\label{sec:random_walk}

We discuss in this section the motion of a suspended particle in an
incompressible fluid of different density, when the fluid is subjected
to an effective acceleration field of the type described in Section
\ref{sec:g-jitter}. This type of motion has also been termed inertial random
walk, because of the similarity with Brownian motion. The difference, of
course, is that the random motion of the particle is not induced by
thermally induced collisions with the molecules of the fluid, but it
results from an effective random buoyancy force acting on the particle.
A qualitative analysis of this process was already given in \cite{re:regel87}.

Consider a spherical particle of radius $R$ and density $\rho_{p}$ 
submerged in an
incompressible fluid of density $\rho_{f}$. If the fluid is enclosed by
perfectly rigid boundaries, the buoyancy force acting on the submerged
particle is $\vec{F}_{b} = \frac{4}{3} \pi 
\left( \rho_{p} - \rho_{f} \right) R^{3} \vec{g}(t)$,
where $\vec{g}(t)$ is the effective acceleration field. In the frame of
reference co-moving with the container enclosing the fluid, $\vec{g}(t)$
is a body force, with intensity equal to the value of the acceleration
of the container. For containers of reasonable size in a microgravity
environment, $\vec{g}$ can be assumed to be spatially uniform.
Viscous friction will act on the particle. Neglecting memory terms and  
corrections due to the finite size of the container, the viscous force
is given by Stokes' formula $\vec{F}_{v} = -6 \pi \eta R \vec{v}$,
where $\eta$ is the shear viscosity of the fluid, and $\vec{v}$ the
velocity of the particle relative to the fluid at infinity. For simplicity, we 
consider in what follows a one dimensional case and write,
\begin{equation}
\label{eq:lang}
\ddot{x} + \gamma \dot{x} = \Delta \rho g(t),
\end{equation}
with $\gamma = 9 \eta / (2 \rho_{p} R^{2})$ and $\Delta \rho = (\rho_{p}
- \rho_{f})/\rho_{p}$. This is a standard Langevin equation which
can be integrated formally to solve for the velocity $v=\dot{x}$.
Squaring the formal solution and taking the ensemble average one finds,
\begin{eqnarray}
\label{eq:v2}
\left< v^{2} \right> & = & \left< v(0)^{2} \right> e^{-2\gamma t} + 2 e^{-2
\gamma t} \Delta \rho \int_{0}^{t} dt' e^{\gamma t'} \left< v(0)
g(t') \right> + \nonumber \\
& & \Delta \rho^{2} e^{-2 \gamma t} \int_{0}^{t} dt'
\int_{0}^{t} dt'' e^{\gamma (t'+t'')} \left< g(t')g(t'')\right>.
\end{eqnarray}
The first two terms account for the statistics of the initial condition 
and are therefore arbitrary. Note that in general, 
the correlation between $v(t)$ and the noise $g(t')$ for $t'>t$ does not 
vanish provided that $g(t)$ has a finite correlation time. 
For monochromatic 
noise ($\tau \rightarrow \infty$), this correlation 
is a non-decaying oscillating function of the time difference $(t-t')$ 
in the limit  $t,t' \rightarrow \infty$. 
Therefore, the second term of 
the right hand side decays at least as $e^{-\gamma t}$ and does not contribute 
for $\gamma t \gg 1$. Two cases of initial conditions are particularly 
simple. First the case when the particle starts at $t=0$ with 
zero velocity and has not been subject to the influence of the noise for 
$t<0$. In this case $\left< v^{2}(0) \right>=0$ and $\left< v(0)g(t) \right>$=0 
and Eq.(\ref{eq:v2}) reduces to 
\begin{eqnarray}
\label{eq:v2t}
\left< v^{2} \right> & = & \frac{\Delta \rho^{2} \left< g^{2} \right>
\tau}{ \left[ \left( \gamma + \lambda \right)^{2} + \Omega^{2} \right]
\left[ \left( \gamma - \lambda \right)^{2} + \Omega^{2} \right]} \times
\nonumber \\
& & \left[ \left(1-e^{-2 \gamma t}\right)\frac{\lambda^{2}}{\gamma} \left(
\lambda^{2} - \gamma^{2} + \Omega^{2} \right) + 
\left(1+e^{-2 \gamma t}\right) \lambda \left( - \lambda^{2} + \gamma^{2}
+ \Omega^{2} \right) \right. \nonumber \\
& & \left. + 2 \lambda \left( \lambda^{2} -
\gamma^{2} - \Omega^{2} \right) e^{-(\gamma + \lambda)t} \cos{\Omega t} -
4 \Omega \lambda^{2} e^{-(\gamma + \lambda)t} \sin{\Omega t} \right].
\end{eqnarray}
The second and physically most relevant choice for the initial condition
is to take precisely the statistics of the steady state. In this case, 
the transient terms of Eq.(\ref{eq:v2}) will cancel exactly the decaying 
terms of Eq. (\ref{eq:v2t}) giving a constant value 
for $\left< v^{2} \right>$  
\begin{equation}
\label{eq:v2final}
\left< v^{2} \right>_{\infty} = \frac{\Delta \rho^{2} \left< g^{2}
\right> \left( \gamma + \frac{1}{\tau} \right)}{\gamma \left[ \left( \gamma +
\frac{1}{\tau} \right)^{2} + \Omega^{2} \right] }.
\end{equation}
This is also the solution for arbitrary initial conditions when
 $\gamma t \gg 1$.

Consider now the limits of white and monochromatic noise. For white
noise, one has,
\begin{equation}
\label{eq:v2white}
\lim_{\tau \rightarrow 0} \left< v^{2} \right>_{\infty} = \frac{\Delta
\rho^{2} \left< g^{2} \right> \tau}{\gamma} = \frac{\Delta
\rho^{2} D}{\gamma},
\end{equation}
and in the monochromatic case,
\begin{equation}
\label{eq:v2mono}
\lim_{\tau \rightarrow \infty} \left< v^{2} \right>_{\infty} =
\frac{\Delta \rho^{2} \left< g^{2} \right>}{\gamma^{2} + \Omega^{2}}.
\end{equation}
(An average over phases of the deterministic forces is assumed in 
the ensemble average of the monochromatic noise limit, otherwise 
$\left< v^ {2} \right>_{\infty}$ would be an oscillatory quantity.)
In the white noise limit, $\left< v^{2} \right>_{\infty}$ is given by a
fluctuation-dissipation relation since $\Delta \rho^{2} D$ is the
intensity of the fluctuations, and $\gamma$ the intensity of the
dissipation. In the
monochromatic limit, however, $\left< v^{2} \right>_{\infty} \propto
1/\gamma^{2}$ (for low frequencies, $\Omega / \gamma \ll 1$). This is
precisely the overdamped limit of Eq. (\ref{eq:lang}). In all
cases the mean
squared value of the velocity saturates at a finite value at long times.
In the monochromatic noise limit, this is the case even in the limit of
small viscosity $\gamma \rightarrow 0$. In the white noise limit, on the
other hand, viscosity is essential for saturation.

The equation for
$\left< x(t)^{2} \right>$ can be obtained by multiplying Eq.
(\ref{eq:lang}) by $x$, and rewriting it as,
\begin{equation}
\label{eq:secmom}
\frac{1}{2} \frac{ d^{2} x^{2}}{dt^{2}} - \left( \frac{dx}{dt}
\right)^{2} + \frac{\gamma}{2} \frac{d x^{2}}{dt} = \Delta \rho x g(t).
\end{equation}
This equation is cumbersome to solve in the general case of narrow band
noise. We focus separately on the two limits of white and monochromatic
noise. In the standard case of white noise, we have $\left< x(t) g(t) \right> = 
0$, and 
$ \left< v(0)g(t>0) \right> =0$ and (from Eq. (\ref{eq:v2})),
\begin{equation}
\left< v^{2} \right> = \frac{\Delta \rho^{2} D}{\gamma} \left( 1 - e^{- 2
\gamma t} \right) + \left<v(0)^2\right> e^{-2 \gamma t}.
\end{equation}
If we assume $\left<v(0)^2\right>$ to take the steady state value 
$\frac{\Delta \rho^{2} D}{\gamma}$ (or equivalently, for an arbitrary initial 
condition if $\gamma t \gg 1$), and taking
the ensemble average of Eq. (\ref{eq:secmom}), we have for $\gamma t \gg 1$ 
$\left< x^{2}(t) \right> = \frac{2 \Delta \rho^{2}
D}{\gamma^{2}} t + {\rm constant}$,
that is diffusive motion with an effective diffusion coefficient
$ \Delta \rho^{2} D /\gamma^{2}$.

For the monochromatic case, the displacement of a particle with $x(0)=0$ 
and $v(0)=0$ can be easily obtained.
Square the deterministic solution for $g(t)=g_0 \cos(\Omega t + \phi)$ and 
average over the phase $\phi$. 
With the identification $\left<g^2\right>=g_0^2/2$ 
and for $\gamma t \gg 1$, we obtain 
\begin{equation}
\label{eq:x2mono}
\left< x^{2}(t) \right> = \frac{\Delta\rho^2\left<g^{2}\right>}
{(\gamma^2+\Omega^2)^2}
\left[1+\left(\frac{\gamma}{\Omega}\right)^2 + 
\left(\frac{\gamma}{\Omega}+\frac{\Omega}{\gamma}
\right)^2
-2\left(\frac{\gamma}{\Omega}+\frac{\Omega}{\gamma}
\right) \left(\frac{\gamma}{\Omega}\cos\Omega t + \sin\Omega t \right)
\right] .
\end{equation}
In this case the mean square displacement of the particle is bounded. 
For the particular case of $\gamma=0$, Eq. (\ref{eq:x2mono}) is not 
valid and the mean square displacement of the particle may be unbounded.

It is worth remarking that Eq. (\ref{eq:x2mono}) is only valid in the strict 
limit $\tau=\infty$. If the correlation time $\tau$ is finite, there is 
always a diffusive behavior superimposed to the oscillations. 
An effective diffusion coefficient $D_{eff}$ 
can be defined as,
\begin{equation}
D_{eff} = \lim_{t \rightarrow \infty} \frac{1}{T} \int_{t}^{t+T} 
\frac{1}{\gamma} \left[
\left< v^{2}(t') \right> + \Delta \rho \left< x(t') g(t') \right>
\right] dt', ~~~~ T = \frac{2 \pi}{\Omega}.
\end{equation}
For narrow band noise we find,
\begin{equation}
\lim_{t\rightarrow \infty} \left<x(t)g(t)\right> = \Delta\rho \left<g^2\right> 
\frac{\lambda(\lambda+\gamma)-\Omega^2}{\left((\lambda+\gamma)^2+
\Omega^2\right)(\lambda^2+\Omega^2)} + {\rm Oscillatory ~ terms},
\end{equation}
and hence
\begin{equation}
\label{eq:deff}
\frac{D_{eff} \gamma}{\Delta\rho^{2} \left<g^2\right>}=
\frac{\frac{\lambda}{\gamma}(\lambda^2-\gamma^2+\Omega^2)+\gamma^2-\lambda^2+
\Omega^2}
{\left( (\gamma+\lambda)^2+\Omega^2\right)\left((\gamma-\lambda)^2+
\Omega^2\right)} + 
\frac{\lambda(\lambda+\gamma)-\Omega^2}
{\left( (\gamma+\lambda)^2+\Omega^2\right)\left(\lambda^2+\Omega^2\right)}.
\end{equation}
As a first correction to the monochromatic limit, 
the effective diffusion coefficient to first order in $1/\tau$ reads  
\begin{equation}
D_{eff}=\frac{\Delta\rho^2 \left<g^2\right>}
{(\gamma^2+\Omega^2)^2} \left(2+\frac{\Omega^2}{\gamma^2}+
\frac{\gamma^2}{\Omega^2} \right)\frac{1}{\tau}+ O(1/\tau^2).
\end{equation}

\section{Coarsening in solid-liquid mixtures}
\label{sec:coarsening}

We consider next coarsening of a solid-liquid mixture
being subjected to a fluctuating acceleration field of the 
type described in section \ref{sec:g-jitter}. Such a study is relevant in
connection with an experiment that will be conducted in
microgravity in the near future. The experiment will address the
asymptotic power law growth governing coarsening, as well as the 
dependence of the amplitude of the growth law on the volume fraction of the 
precipitate phase. The absence of gravitationally induced sedimentation
will allow a careful quantitative study of these two important
theoretical issues.
%
%

A residual acceleration field can produce a number of deleterious
effects on otherwise purely diffusive controlled coarsening, which we 
address in this
section. We focus here on two such effects: random motion of the suspended
particles induced by the effective (random) buoyant force and the concomitant
increase in the likelihood of particle coalescence, and additional flow in the 
fluid phase caused by g-jitter and its effect on solute mass transport.

In order to obtain numerical estimates for these two effects, we will
consider experimental parameters for a solid-liquid mixture of Sn-rich 
particles in a Pb-Sn eutectic liquid, the system that will be used in the 
microgravity experiment \cite{re:hardy88,re:hardy91}. 
The density of the precipitating solid phase is 
$\rho_{p} = 7.088 g/cm^{3}$,
whereas that of the liquid is $\rho_{l} = 8.074 g/cm^{3}$. The kinematic 
viscosity of the liquid is $\nu = 2.48 \times 10^{-3} cm^{2}/s$, and the
solute diffusivity is $D_{s} = 5.6 \times 10^{-6} cm^{2}/s$. It is
anticipated that coarsening will be studied for a period of 5 hours,
with an average particle size at the end of that period of
$R_{av} \simeq 7 \times 10^{-3} cm = 70 ~ \mu m$. Given the size of the
particles and the small values of the residual gravitational field,
inertial effects will be completely negligible. In what follows, we 
focus almost exclusively on Stokesian dynamics for the suspended
particles \cite{fo:ross2_3}. In
addition, the solution is not mono-disperse, but rather a scale invariant
particle size distribution evolves dynamically 
\cite{re:gunton83}, with particles sizes ranging from 0 to
$\approx 1.5 R_{av}$. In all the estimates that follow, the average size at
the end of the five hours is used, an overestimate for most of the
duration of the experiment, and a slight underestimate at the latest
times. As will be seen below, a factor of two in $R_{av}$ would not modify
our conclusions.

Neglecting inter-particle hydrodynamic interactions, precipitate particles 
will execute a 
random motion of the type described in Section \ref{sec:random_walk}. 
For the case of monochromatic noise (fixed frequency and random phase), 
the average
quadratic displacement of each particle remains bounded, and is given by
Eq. (\ref{eq:x2mono}). For the parameters of the fluid given
$\gamma = 260 s^{-1}$, and by using the amplitude of the 17 Hz component 
of the power spectrum in Section \ref{sec:g-jitter} ($\Omega = 2 \pi
~ 17 s^{-1}, <~g^{2} > = 1.27 \times 10^{-7} g_{E}^{2}$), we find that
max $ \{ <x^{2}>  \} \approx 10^{-8} cm^{2}$, and hence negligible. 
At the other
extreme, we find that for white noise (replacing the
factor of 2 by 6 corresponding to diffusion in three dimensional space
in the expression for the effective diffusion coefficient),
the mean squared displacement after five hours is
$ <x^{2}> (t = {\rm 5 ~ hr.}) = 8.85 \times 10^{-6} cm^{2}$ or
$\sqrt{<x^{2}>} \simeq 30 \mu m $.
Clearly the average square displacement induced by the white noise
component of the residual acceleration
field is much larger than that induced by the monochromatic
component, but it is still about one half of the average particle size. 
Therefore random 
motion of particles induced by g-jitter will not lead to 
significant coalescence
during this time period. As a reference, 
we quote the average
squared displacement induced by thermal Brownian motion:
$ \left< x^{2} \right> = k_{B}Tt/3 \pi R_{av} \eta $
or $\sqrt{\left< x^{2} \right> } \simeq 39~\mu m$,
for the same time span and taking $R_{av} = 70~\mu m$. Both
effects are therefore expected to be of the same order of magnitude during
the experiment.

It is also possible to estimate hydrodynamic interaction effects between 
pairs of spherical particles to show that it leads to weak attraction at long
distances, and repulsion at short distances. The
relative displacement $\vec{r}$ of particle 2 with respect to particle 1
immersed in an incompressible
fluid satisfies \cite{re:batchelor76,re:russel89},
\begin{equation}
\frac{d \vec{r}}{dt} =  \left( \mbox{\boldmath{$\omega$}}_{21} - 
\mbox{\boldmath{$\omega$}}_{11}
\right) \cdot \vec{F}_{1} +
 \left( \mbox{\boldmath{$\omega$}}_{22} - \mbox{\boldmath{$\omega$}}_{21}
\right) \cdot \vec{F}_{2},
\end{equation}
where $\vec{F}_{i}$ is the force acting on the $i$-th particle,
and \mbox{\boldmath{$\omega$}}$_{ij}$ are hydrodynamic mobility
tensors, given, e.g., in references \cite{re:batchelor76,re:russel89}.
After some straightforward algebra, the leading contribution at distances
large compared to the particle radii is given by,
\begin{equation}
\label{eq:interaction}
\frac{d \vec{r}}{dt} = \frac{2 (\rho_{p}-\rho_{f})}{ 9 \mu}
\left( R_{2}^{2} - R_{1}^{2} \right) \vec{g}(t) + 
\frac{ (\rho_{p}-\rho_{f}) \left( R_{1}^{3} - R_{2}^{3}
\right) }{3 \mu} \frac{1}{r} \left[ \frac{\vec{r} \vec{r}}
{r^{2}} + \frac{1}{2} \left( {\cal I} - \frac{\vec{r} \vec{r}}
{r^{2}} \right) \right] \cdot \vec{g}(t),
\end{equation}
where ${\cal I}$ is the identity tensor.
The first term in the right hand side describes the relative motion of
two independent particles of different size, and therefore its magnitude
has already
been estimated above. Both the longitudinal and transverse components of
the second term in the right hand side of Eq. (\ref{eq:interaction}) are of 
the form,
\begin{equation}
\label{eq:rmodel}
\frac{dr}{dt} = \frac{A}{r} g(t),
\end{equation}
where, for the longitudinal component, $A=(\rho_{p} - \rho_{f}) 
(R_{1}^{3} - R_{2}^{3} )/3 \mu$.

Consider an initial inter-particle separation $r_{0} \gg R_{i}$.
In this case, and for times shorter than the average time needed for the two
particles to coalesce,
the quantity $y = r^{2}/2A$ is a Wiener process if $g(t)$ is Gaussian and 
white, and therefore the conditional probability for $r$ is,
\begin{equation}
P(r, t | r_{0}, t_{0}) = \frac{r}{|A| \sqrt{4 \pi D (t-t_{0})}} 
e^{- \frac{ \left( r^{2} -r_{0}^{2} \right)^{2}}{ 16 D A^{2} (t-t_{0})}}.
\end{equation}
The ensemble average of $r$, $<r>$ can be computed analytically,
\begin{equation}
<r> = \sqrt{ \frac{|A|}{4}} \left[ 2D (t-t_{0}) \right] e^{- 
\frac{r_{0}^{4}}{32 A^{2} D (t-t_{0})}} D_{-3/2} \left(
- \frac{r_{0}^{2}}{ 2|A| \sqrt{2D (t-t_{0})}} \right),
\end{equation}
where $D_{p}(z)$ is a parabolic cylinder function \cite{re:gradshteyn80} 
(formula 9.240). For short times, the asymptotic form of $D_{p}(z)$ 
for large $z$ allows the computation of $<r>$
\begin{equation}
\label{eq:rav}
 <r> = r_{0} \left( 1 - \frac{A^{2} D (t-t_{0})}{r_{0}^{4}} \right),
\end{equation}
which decreases in time regardless of the sign of $A$. Therefore 
g-jitter induces an effective hydrodynamic attraction between pairs of
particles. However, for the experimental values given above, and taking
$R_{1} = 1.5 R_{av}$ and $R_{2} = 0.5 R_{av}$, $A = 1.8 \times 10^{-5}
cm ~s$. If $r_{0} \simeq 200 \mu m$, then after 5 hours the 
inter-particle separation would have decreased by approximately 
$7 \mu m$, and therefore small compared to particle radii.

The attractive interaction is not confined to short times, but it
arises directly from the $1/r$ dependence in Eq. (\ref{eq:rmodel}).
By taking the average of Eq. (\ref{eq:rmodel}), using
the Furutsu-Novikov theorem \cite{re:hanggi85} and the fact that the noise is 
Gaussian and white, one finds,
\begin{equation}
\frac{ d<r> }{dt} = A D \left< \frac{ \delta 1/r(t) }{ \delta g(t) } \right>,
\end{equation}
where $\delta / \delta g(t)$ stands for functional derivative with
respect to $g$.  Directly from Eq. (\ref{eq:rmodel}), we find that
$ \delta (1/r(t)) / \delta g(t) = -A/r^{3}$, and therefore,
\begin{equation}
\frac {d <r> }{dt} = -A^{2} D \left< \frac{1}{r^{3}} \right>,
\end{equation}
identical to Eq. (\ref{eq:rav}) with $1/r_{0}^{3}$ replaced by
$< 1/r^{3} >$. Since $r$ is a positive quantity, $d<r>/dt < 0$
for all values of $r$. It is also interesting to note that the
effective attractive interaction is not confined to the 
term proportional to $1/r$ in the hydrodynamic mobility,
but that attractive contributions arise from higher powers of $1/r$ as 
well. In fact, this attraction is generic for overdamped motion and
multiplicative noise provided that
the mobility is a decaying function of the inter-particle separation
\cite{fo:ross2_2}.

The question naturally arises as to the behavior of pairs of particles
near contact, or of particles near a solid wall. In either case,
lubrication theory allows the calculation of the mobility tensor.
The longitudinal component vanishes linearly with inter-particle 
distance whereas the transverse component becomes
non-analytic (diverges logarithmically at short distances)
\cite{re:russel89}.
In both cases, the mobility {\em increases} with inter-particle 
separation leading to an average repulsion ($d<r>/dt > 0$)
following the same arguments given above.

Estimating the effect of g-jitter on mass transport in the fluid phase
and therefore on coarsening kinetics
is far more complex, and we will not attempt a complete solution here.
We show below that the order of magnitude of this contribution to
coarsening is also small, and therefore a detailed calculation is not
necessary. However, and in order to motivate the analysis that
follows, let us define an effective Peclet number as
$ {\rm Pe} = \sqrt{<u^{2}>} R / D_{s}$,
where $\vec{u}$ is the characteristic velocity of the fluid. (The 
velocity $\vec{u}$ is of the order of the velocity of the particles
because the
motion of two-phase interfaces due to phase change is small in
the time scale of change of $g(t)$). We find for monochromatic noise 
(Eq. (\ref{eq:v2mono})) that $<u^{2}> = 3.48 \times 10^{-8} cm^{2}/s^{2}$ 
or ${\rm Pe} = 0.23$. For white noise, on the other hand, according to Eq. 
(\ref{eq:v2white}),
$<u^{2}> = 6.39 \times 10^{-8} cm^{2}/s^{2}$, or ${\rm Pe} = 0.32$. 
Therefore it would appear that convective transport of mass is not
negligible in front of diffusive transport.

Such a calculation, however, overestimates convective transport.
Since the system is statistically uniform and the effective 
acceleration field averages to zero, the average velocity of the fluid has 
to be zero. Hence no overall convective motion would
result in the (longer) time scale over which diffusive transport occurs,
once the fluctuating component is averaged over times much larger than
the correlation time of the noise. Since $<\vec{u}> = 0$
but $<u^{2}> \neq 0$, the lowest order contribution to transport
due to the motion of a fluid element is diffusive \cite{fo:ross2_1}.
The calculation of this effective diffusivity can be carried out in a
mean field approximation. Consider a single solid particle immersed
in the fluid phase, so that the presence of the remaining particles can
be subsumed in a far field composition $c_{\infty}$ 
\cite{re:gunton83}, and a far field velocity $\vec{u}_{\infty}$,
both to be determined self-consistently for a given particle 
distribution. The cut-off distance is typically of the order of the 
inter-particle separation.

Fluctuations in $\vec{u}_{\infty}$ due to the motion of the
ensemble of particles (again, fast
compared to coarsening times) lead to an effective increase in diffusive
mass transport, and hence to an increased diffusivity. For a quiescent system,
the average velocity of the fluid $\vec{u}_{av}$, and the average 
velocity of the particles $\vec{v}_{av}$ are related by,
\begin{equation}
\phi \vec{v}_{av} + ( 1 - \phi ) \vec{u}_{av} = 0,
\end{equation}
where $\phi$ is the volume fraction of the system, which we assume to be
small. The subindex {\em av} indicates an average over
the particle distribution at fixed time. In mean field, we take
$\vec{u}_{\infty} = \vec{u}_{av}$, and also
\begin{equation}
\vec{v}_{av} = \frac{ 2 R_{av}^{2} ( \rho_{p} - \rho_{f} ) \vec{g}(t) }
{9 \eta} + {\cal O} (\phi),
\end{equation}
with hydrodynamic interactions contributing to ${\cal O}(\phi)$. Therefore,
\begin{equation}
\vec{u}_{\infty} = - \frac{ 2 \phi R_{av}^{2} ( \rho_{p} - \rho_{f} ) 
\vec{g}(t) }{ 9 \eta} + {\cal O} (\phi^{2}).
\end{equation}
The far field effective diffusivity is then \cite{re:boon91}
\begin{equation}
D_{eff} = D_{s} + \frac{1}{3} \int_{0}^{\infty} \left< \vec{u}_{\infty} (t) \cdot
\vec{u}_{\infty}(t+t') \right> dt' = D_{s} + \frac{\Delta \rho^{2} \phi^{2} D}
{\gamma_{av}^{2}}
\end{equation}
in the white noise limit. For the parameters of the experiment given above
$$
\frac{\Delta \rho^{2} \phi^{2} D}{\gamma_{av}^{2} D_{s}} \ll 1,
$$
and therefore numerically negligible.

In summary, even though anticipated Peclet numbers based on the scale of the
flow are of order unity, mass transport due to convection is expected to be
negligible during the solid-liquid coarsening experiment. Since the time
scale of acceleration variations is short compared to coarsening times, and
leads to zero average velocity, the contribution from g-jitter to mass 
transport is diffusive and leads to a very small correction to the solute
diffusivity. We note, however, that the correction is proportional to
$1/\gamma_{av}^{2} \propto R_{av}^{4}$ and hence it increases
quickly with the average particle size of the precipitate phase.
Therefore, either under different experimental conditions, or in the
strict asymptotic limit of very long times (and hence large $R_{av}$),
transport due to transient accelerations would dominate molecular diffusion
leading to a different asymptotic growth law for the average particle size.

\section*{Acknowledgments}

We are indebted to Richard DeLombard and the PIMS project at Lewis
Research Center for providing us with the SAMS accelerometer data used in
this paper, and also to Peter Voorhees for information on the experiment
on \lq\lq Coarsening in Solid-Liquid Mixtures". 
This work is supported by the Microgravity Science and Applications
Division of the NASA under contract No. NAG3-1284,
and also in part by the Supercomputer
Computations Research Institute, which is partially funded by the U.S.
Department of Energy, contract No. DE-FC05-85ER25000.
JC is also supported by the Direcci\'on General de Investigaci\'on 
Cient\'\i fica y T\'ecnica, contract No. PB93-0769.

\bibliographystyle{physfla}
\bibliography{references}

\begin{thebibliography}{10}
\bibitem{fo:ross2_3}
Inertial effects must be included for the case of white noise, or, in general,
  for colored noise when the correlation time $\tau$ is smaller or of the order
  of the viscous damping time $1/\gamma$ (see Eq. (\ref{eq:v2final})). In the
  case considered, $1/\gamma = 1/260~s$, smaller than the sampling period of
  the acceleration data. Second, Stokes dynamics also fails at distances longer
  than the momentum diffusion length $l_{\nu}$. In the case at hand $l_{\nu} =
  \nu / u \simeq (2.48\times 10^{-3}~cm^{2}/s)/(10^{-4}~cm/s) = 24.8~cm$ larger
  than the dimensions of the container. Hence the motion of the fluid
  surrounding the particles is overdamped, and satisfies the equation for
  Stokes flow driven by the instantaneous velocity of the particles.
\bibitem{fo:ross2_2}
It is important to emphasize the relevance of overdamped motion to the results
  presented. If the correlation time of the fluctuating force is smaller than
  the viscous decay time (as is the case in thermal Brownian motion), the
  fluctuating force enters additively in the equations of motion for the
  particles and the effects discussed do not arise. In the present case,
  however, the viscous decay time is $1/260 ~s$, much smaller than the
  correlation time of the effective acceleration field. In this limit, the
  fluctuating component enters multiplicatively and leads to an average
  attraction at long distances (but still shorter than the visous decay
  length), and repulsion at short distances.
\bibitem{fo:ross2_1}
There is no large scale (low wavenumber) motion as in the case of turbulent
  flow. The Reynolds number is typically much smaller than unity.
\end{thebibliography}

\end{document}